\documentclass[conference]{IEEEtran}
\usepackage{amssymb}
\usepackage{amsmath}

\usepackage{graphicx}
\usepackage{caption}
\usepackage{subcaption}

\ifCLASSINFOpdf
\else
\fi
\begin{document}
%
\title{TrustConnect: An In-Vehicle Anomaly Detection Framework through Topology-Based Trust Rating}

\author{\IEEEauthorblockN{Ayan Roy and Jeetkumar Patel}
\IEEEauthorblockA{School of Engineering and Computing\\
Christopher Newport University\\
Newport News, Virginia\\
Email: \{ayan.roy, jeetkumar.patel.23\}@cnu.edu}
\and
\IEEEauthorblockN{Rik Chakraborti}
\IEEEauthorblockA{Department of Economics\\
Christopher Newport University\\
Newport News, Virginia\\
Email: rik.chakraborti@cnu.edu}
\and
\IEEEauthorblockN{Shudip Datta}
\IEEEauthorblockA{ Software Engineer \\
Email: shudipdatta@mst.edu}}

%


\maketitle

\begin{abstract}
Modern vehicles are equipped with numerous in-vehicle components that interact with the external environment through remote communications and services, such as Bluetooth and vehicle-to-infrastructure communication. These components form a network, exchanging information to ensure the proper functioning of the vehicle. However, the presence of false or fabricated information can disrupt the vehicle’s performance. Given that these components are interconnected, erroneous data can propagate throughout the network, potentially affecting other components and leading to catastrophic consequences. To address this issue, we propose TrustConnect, a framework designed to assess the trustworthiness of a vehicle's in-vehicle network by evaluating the trust levels of individual components under various network configurations. The proposed framework leverages the interdependency of all the vehicle’s components, along with the correlation of their values, and the vulnerability to remote injection based on the outside exposure of each component, to determine the reliability of the in-vehicle network. The effectiveness of the proposed framework has been validated through programming simulations conducted across various scenarios, using a random distribution of an in-vehicle network graph generated with the Networkx package in Python.
\end{abstract}
\begin{IEEEkeywords}
   Trust Management,
   In-Vehicle Network,
   CAN,
   Anomaly Detection,
   Security
\end{IEEEkeywords}
\section{Introduction}

Modern vehicles today are equipped with numerous electronic control units (ECUs) that manage and execute a wide range of functions critical to vehicle operation \cite{walrand2021architecture}. These ECUs oversee essential systems such as autonomous driving, infotainment, safety features, and driver assistance, which significantly contribute to the overall comfort and safety of passengers. The ECUs within a vehicle form an intricate network, often referred to as the \emph{in-vehicle network}, which facilitates communication and information exchange via protocols such as the Controller Area Network (CAN) bus, a defacto
standard for in-vehicle communication owing to its stability
and cost-effectiveness\cite{song2021self}.

Many ECUs are exposed to remote connectivity via Bluetooth, 5G, and other internet services, which increases the risk of remote injection attacks by malicious actors \cite{nakamura2021vehicle}\cite{9922025}. These attacks exploit vulnerabilities in the vehicle's systems \cite{rathore2022vehicle}, with the potential for significant impact, as demonstrated by several researchers \cite{checkoway2011comprehensive}\cite{miller2013adventures}. A report from 2020 revealed several vulnerabilities that could potentially allow for the remote control of Mercedes-Benz vehicles \cite{yan2020security}. A major factor contributing to the success of remote injection attacks is the lack of an authentication mechanism for the information transmitted through the CAN bus \cite{sun2021anomaly}. Given the widespread connectivity of modern vehicles, which include numerous sensors, actuators, and ECUs, ensuring the reliability of data from each component is critical. A fault or security breach in any single ECU can propagate across the network, potentially affecting other components and leading to catastrophic failures. As vehicles increasingly depend on complex, interconnected systems, it is essential to assess and preserve the integrity and trustworthiness of the data exchanged within these networks to prevent such vulnerabilities from compromising vehicle safety and functionality.

In this direction, researchers have invested considerable effort in developing frameworks to detect intrusions or anomalies within in-vehicle networks. Common approaches include machine/deep learning models \cite{wang2024intrusion}, signature-based/sequence systems \cite{marchetti2017anomaly}\cite{wang2024traffic}, frequency/time-based systems \cite{7420322}, all designed to filter out anomalous data or detect intrusions. These methods can also be utilized to create a framework that assesses the trustworthiness of various ECUs within the network. However, each of these approaches has its limitations. One major drawback of machine learning-based models is their high resource consumption, which is often not feasible for in-vehicle networks. Signature-based systems face the issue that an ECU has a fixed set of outputs, but the timing of those outputs is not considered, making them vulnerable to replay attacks by an attacker during a remote injection. Similarly, frequency and time based systems focus on the timing intervals at which an ECU generates messages, but an attacker can exploit these intervals to deceive the system.

In this paper, we introduce TrustConnect, a framework designed to assess the trustworthiness of an in-vehicle network by evaluating the trust of individual ECUs (Electronic Control Units). The ECUs are represented as a directed graph, enabling the evaluation of each ECU’s trust based on its relationships with other ECUs in the network. Furthermore, we account for the potential risks associated with remote exposure, considering the impact of remote injection attacks that could undermine the trust of individual ECUs and compromise the overall integrity of the network. The key contributions of this paper are as follows:

\begin{itemize}
   \item Proposes a novel trust framework for determining the trustworthiness of an in-vehicle network by calculating the individual trust levels of various ECUs within the network. The proposed model does not rely on predefined rules or signatures to identify anomalous data within the ECUs. Instead, the trust of the network is based on a snapshot of the data generated by different ECUs at a specific point in time.
\end{itemize}

\section{Related Works}
A significant amount of research has been dedicated to filtering out anomalous data by designing Intrusion Detection Systems (IDS) based on various strategies. Jin et al. proposed a signature-based IDS \cite{9401087}, which utilizes factors such as message ID, time, correlation, context, and value range. Although the correlation approach considers the relationships between different components, it does not account for the relative weight or importance of each component. This limitation allows an attacker to exploit a more vulnerable system to deceive a defender into thinking that a less vulnerable system has been compromised. This could occur when the correlation of data from a less vulnerable ECU is used with data from a more vulnerable ECU, misleading the detection system.

A similar correlation-based anomaly detection system was proposed by Xiao et al. \cite{xiao2023robust}, which leverages Graph Neural Networks. Their system effectively uses the correlation between changes in traffic bytes and the state of other traffic bytes to identify anomalies. However, like the earlier approach, this system does not consider the varying vulnerability of each component, potentially limiting its effectiveness in certain scenarios.

Tahsin et al. proposed an anomaly detection system \cite{donmez2021anomaly} that detects anomalous data based on the sequence of messages within an in-vehicle network. While this method is effective in certain contexts, it struggles in the case of multi-point injection attacks. In such cases, attackers can inject messages that mimic data from multiple ECUs simultaneously, preserving the sequence while inserting fabricated data, thus evading detection.

Unlike some commonly used strategies, our proposed framework, TrustConnect, leverages the interdependency of the different ECUs within the in-vehicle network, along with the vulnerability factor of each ECU, to assess the overall robustness of the entire network. TrustConnect enhances the system’s resilience against attacks where an attacker targets the most vulnerable component, propagating the attack to make a less vulnerable component appear compromised. Furthermore, the framework places significant emphasis on the outside exposure of each ECU to determine its trustworthiness. This means that even if a strongly connected ECU with minimal exposure is linked to vulnerable ECUs within the network, it is not automatically flagged as anomalous due to injected data in its components.
\section{Preliminaries and Assumptions}
In our proposed model, we assume that certain ECUs have greater exposure to the outside environment than others. The more remote the connection to the outside world, the higher the likelihood of a remote injection attack. This likelihood is represented by a value, denoted as $\epsilon\textsubscript{i}$, for a given ECU $\mathcal{E\textsubscript{i}}$. A higher value of $\epsilon\textsubscript{i}$ indicates a stronger connection to the outside world, thereby increasing the vulnerability to a remote injection attack. 
\section{Proposed Framework}
In the proposed framework, the in-vehicle network, consisting of various ECUs, is represented as a directed graph as shown in Fig \ref{fig:graph}. This graph is an ordered pair, denoted as $\mathcal{(V,E)}$, where $\mathcal{V}$ is the set of all ECUs within the in-vehicle network, and $\mathcal{E}$   set of edges connecting any two ECUs,  $\mathcal{E\textsubscript{i}}$ and $\mathcal{E\textsubscript{j}}$. An edge, $\mathcal{E\textsubscript{i}} \to \mathcal{E\textsubscript{j}}$ signifies that the value of $\mathcal{E\textsubscript{i}}$ can be inferred by $\mathcal{E\textsubscript{j}}$; in other words, $\mathcal{E\textsubscript{i}}$ is dependent on the value of $\mathcal{E\textsubscript{j}}$. Subsequently, we utilize these dependencies among the ECUs within the in-vehicle network to compute the trust score for each ECU, as well as the overall trust score for the entire network.
The weight parameter,  
\begin{figure}
    \centering
    \includegraphics[width=0.5\textwidth]{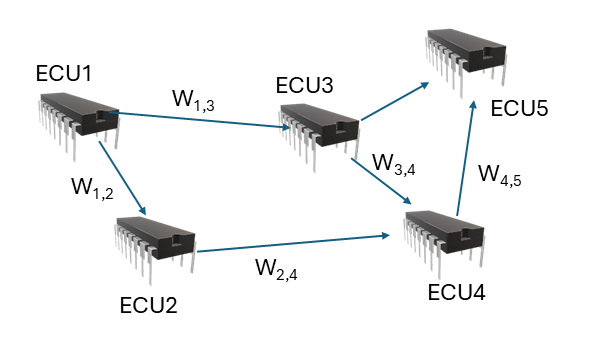}
    \caption{In-Vehicle ECU Dependency Graph: Each ECU is dependent on other ECUs, with a corresponding weight representing the strength of the dependency. }
    \label{fig:ecu_invehicle}
\end{figure}  
\subsection{Trust Influence of One ECU on Another ECU}
For any edge $\mathcal{E\textsubscript{i}} \to \mathcal{E\textsubscript{j}}$, the weight of $\mathcal{E\textsubscript{j}}$ for $\mathcal{E\textsubscript{i}}$ is given in eqn \ref{Eqn: Value inference} and \ref{Eqn: Weight} respectively.
\begin{equation}
\mathcal{D\textsubscript{i,j}}=abs(Val(E\textsubscript{i,i})-Val(E\textsubscript{i,j}))
\label{Eqn: Value inference}
\end{equation}
\begin{equation}
\mathcal{W\textsubscript{i,j}}= e^{-k\cdot \mathcal{D\textsubscript{i,j}}}
\label{Eqn: Weight}
\end{equation}

As per eqn 1, $\mathcal{D\textsubscript{i,j}}$ represents the absolute difference between the value observed from $\mathcal{E\textsubscript{i}}$ and the value that $\mathcal{E\textsubscript{j}}$ calculates for  $\mathcal{E\textsubscript{i}}$ should have at a given point in time. The primary objective of $\mathcal{D\textsubscript{i,j}}$ is to quantify the deviation in the reading between a given ECU,  $\mathcal{E\textsubscript{i}}$,   its neighboring ECU, $\mathcal{E\textsubscript{j}}$ on which $\mathcal{E\textsubscript{i}}$ depends. $k$ is considered a decreasing factor that reduces the weight of $\mathcal{E\textsubscript{j}}$ on $\mathcal{E\textsubscript{i}}$ based on the deviation of the values, as calculated in eqn \ref{Eqn: Value inference}.

The weight parameter, $\mathcal{W\textsubscript{i,j}}$, represents the variation in the values computed by $\mathcal{E\textsubscript{j}}$ for $\mathcal{E\textsubscript{i}}$ when the edge $\mathcal{E\textsubscript{i}} \to \mathcal{E\textsubscript{j}}$ exists. The role of $\mathcal{W\textsubscript{i,j}}$ is to enhance the confidence in $\mathcal{E\textsubscript{i}}$ relative to all other $\mathcal{E\textsubscript{j}}$s, while also amplifying any deviations that may exist. These deviations may arise when $\mathcal{E\textsubscript{j}}$ calculates $\mathcal{E\textsubscript{i}}$ based on its inferred values, further influencing the weight parameter's impact.

\subsection{Calculating individual ECUs trust and Overall Network trust}

The trust score of each $\mathcal{E\textsubscript{i}}$, denoted as $\mathcal{T\textsubscript{E(i)}}$, is computed using equation \ref{trust_ecu}. According to this equation, for every $\mathcal{E\textsubscript{j}}$ where $\mathcal{E\textsubscript{i}} \to \mathcal{E\textsubscript{j}}$ exists, the value of $\epsilon\textsubscript{j}$ is incorporated along with a constant, $\alpha$, which gives greater weight to the likelihood of remote injection. Additionally, the term $\mathcal{C(E\textsubscript{j})}$ is included in the equation. When a trust score for $\mathcal{E\textsubscript{j}}$ is available, $\mathcal{C(E\textsubscript{j})}$ is replaced by $\mathcal{T(E\textsubscript{j})}$. Consequently, the trust score $\mathcal{T(E\textsubscript{i})}$ for $\mathcal{E\textsubscript{i}}$ is calculated using the weight determined in equation \ref{Eqn: Weight} for each $\mathcal{E\textsubscript{i}} \to \mathcal{E\textsubscript{j}}$ relationship.
\begin{equation}
 \mathcal{T(E\textsubscript{i})}=\sum_{j=1}^{n} (\epsilon\textsubscript{j} * \alpha * \mathcal{C(E\textsubscript{j})} + \mathcal{W\textsubscript{i,j}} ) 
\label{trust_ecu}
\end{equation}


\begin{figure}[!hptb]
    \centering
    \includegraphics[width=0.5\textwidth]{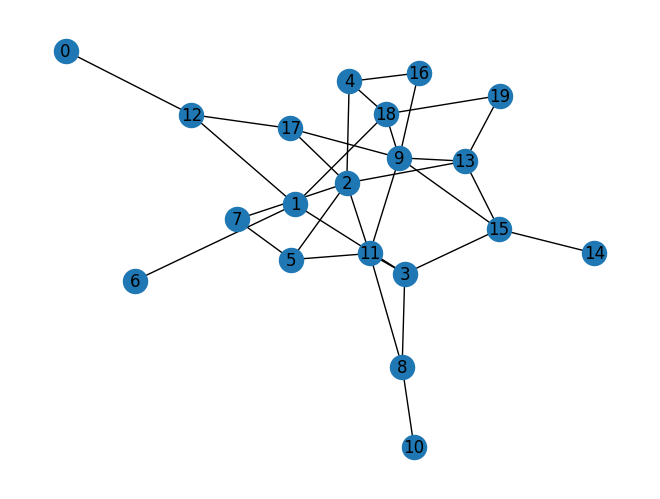}
    \caption{Simulated In-Vehicle Network Dependency Graph}
    \label{fig:graph}
\end{figure}  
\begin{figure}
    \centering
    \includegraphics[width=0.5\textwidth]{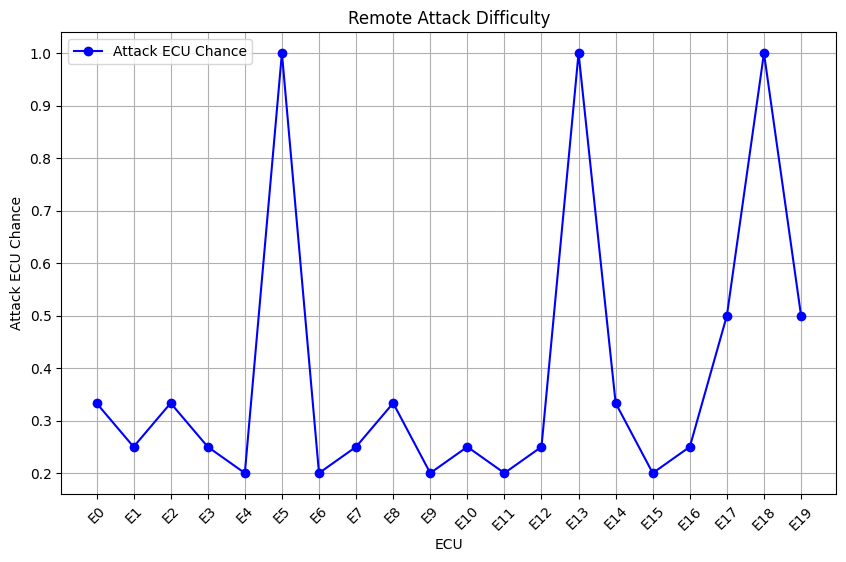}
    \caption{Simulated Remote Attack Difficulty of Individual ECU- 1: Low chance of attack, 0: high chance of attack}
    \label{attack difficulty}
\end{figure}  
\begin{figure*}[!htbp]
\centering
\begin{tabular}{cccc}
\begin{subfigure}{0.23\textwidth}
  \centering
  \includegraphics[width=\linewidth]{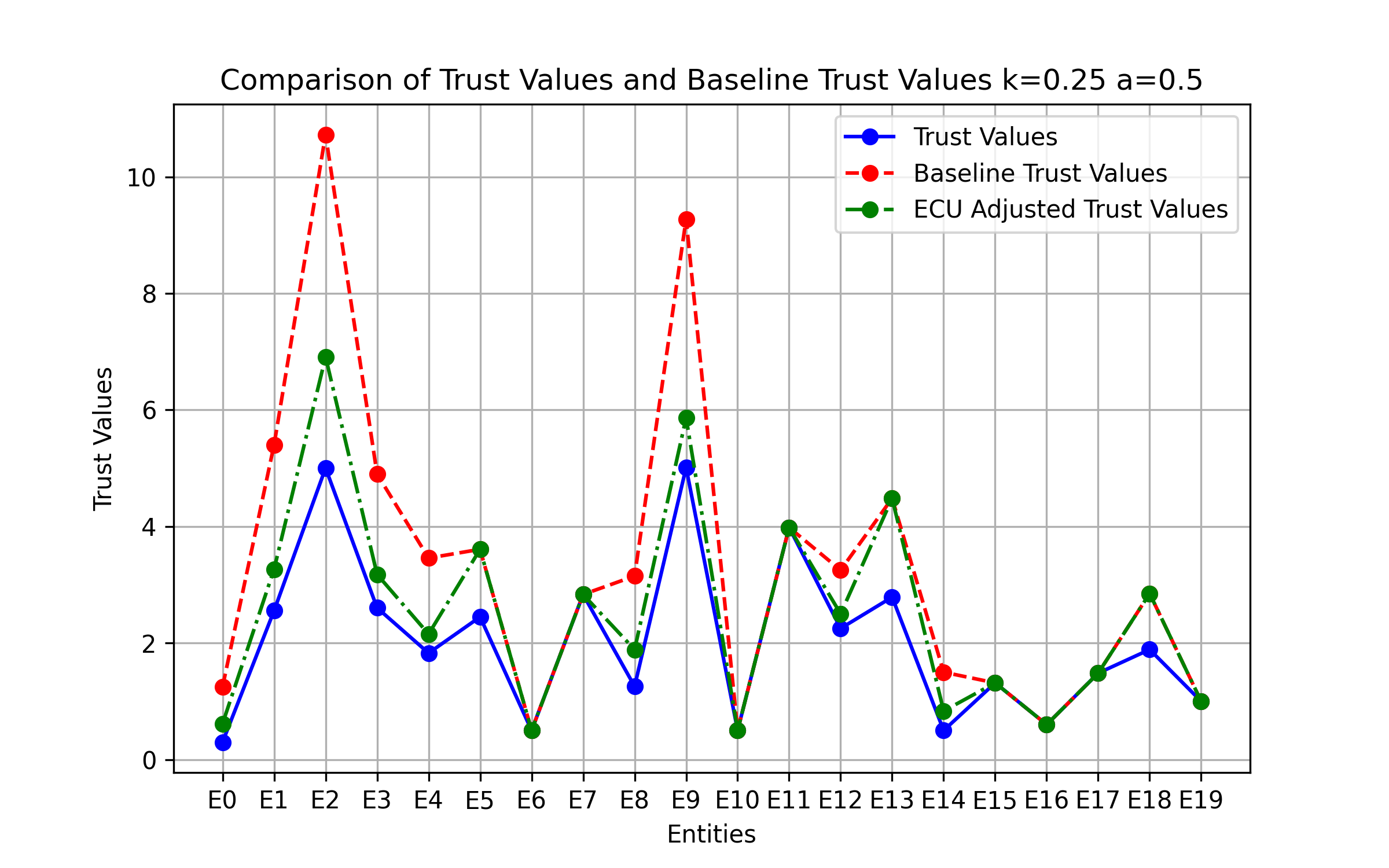}
  \caption{}
  \label{fig:1}
\end{subfigure} &
\begin{subfigure}{0.23\textwidth}
  \centering
  \includegraphics[width=\linewidth]{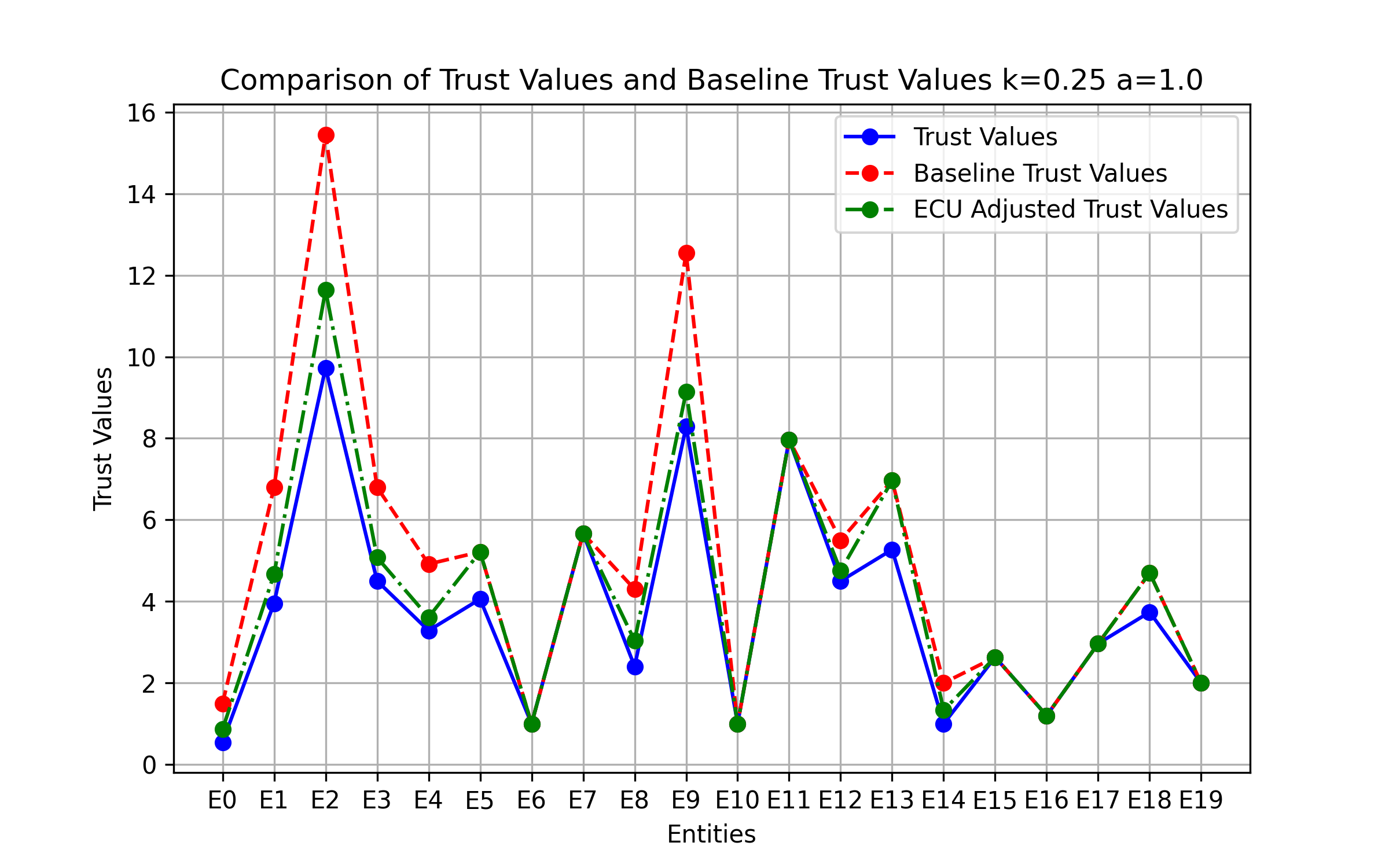}
  \caption{}
  \label{fig:2}
\end{subfigure} &
\begin{subfigure}{0.23\textwidth}
  \centering
  \includegraphics[width=\linewidth]{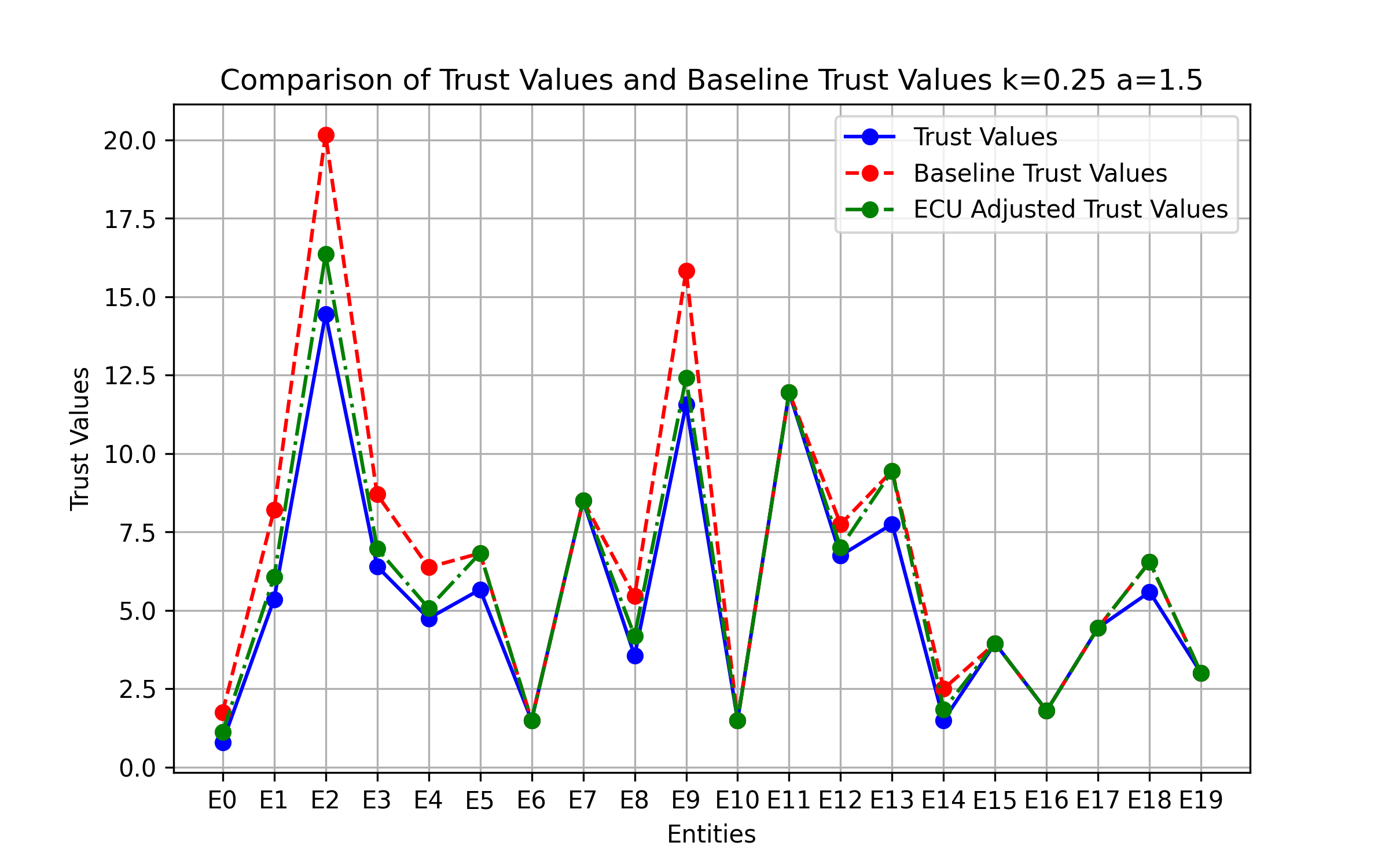}
  \caption{}
  \label{fig:3}
\end{subfigure} &
\begin{subfigure}{0.23\textwidth}
  \centering
  \includegraphics[width=\linewidth]{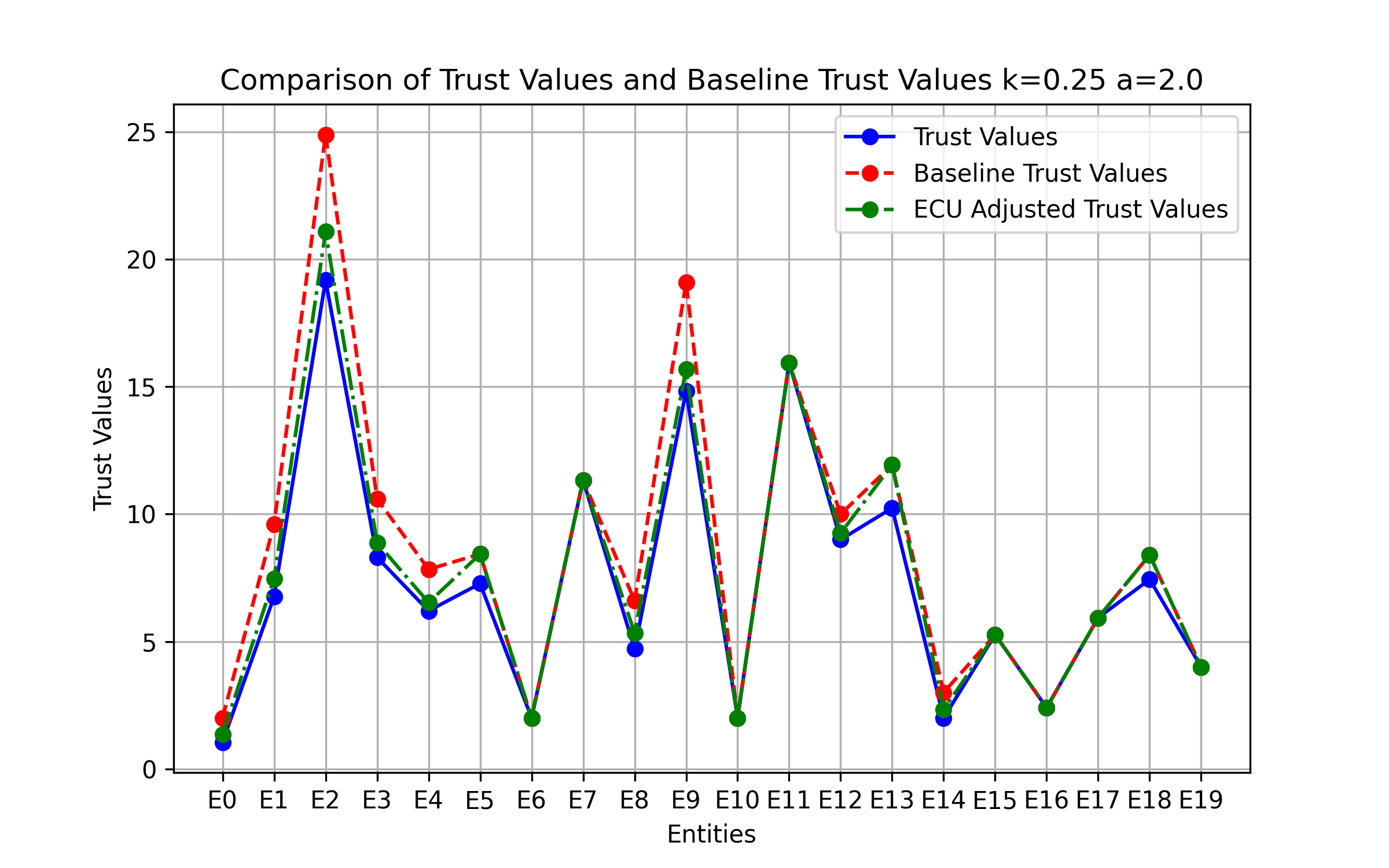}
  \caption{}
  \label{fig:4}
\end{subfigure} \\
\\
\begin{subfigure}{0.23\textwidth}
  \centering
  \includegraphics[width=\linewidth]{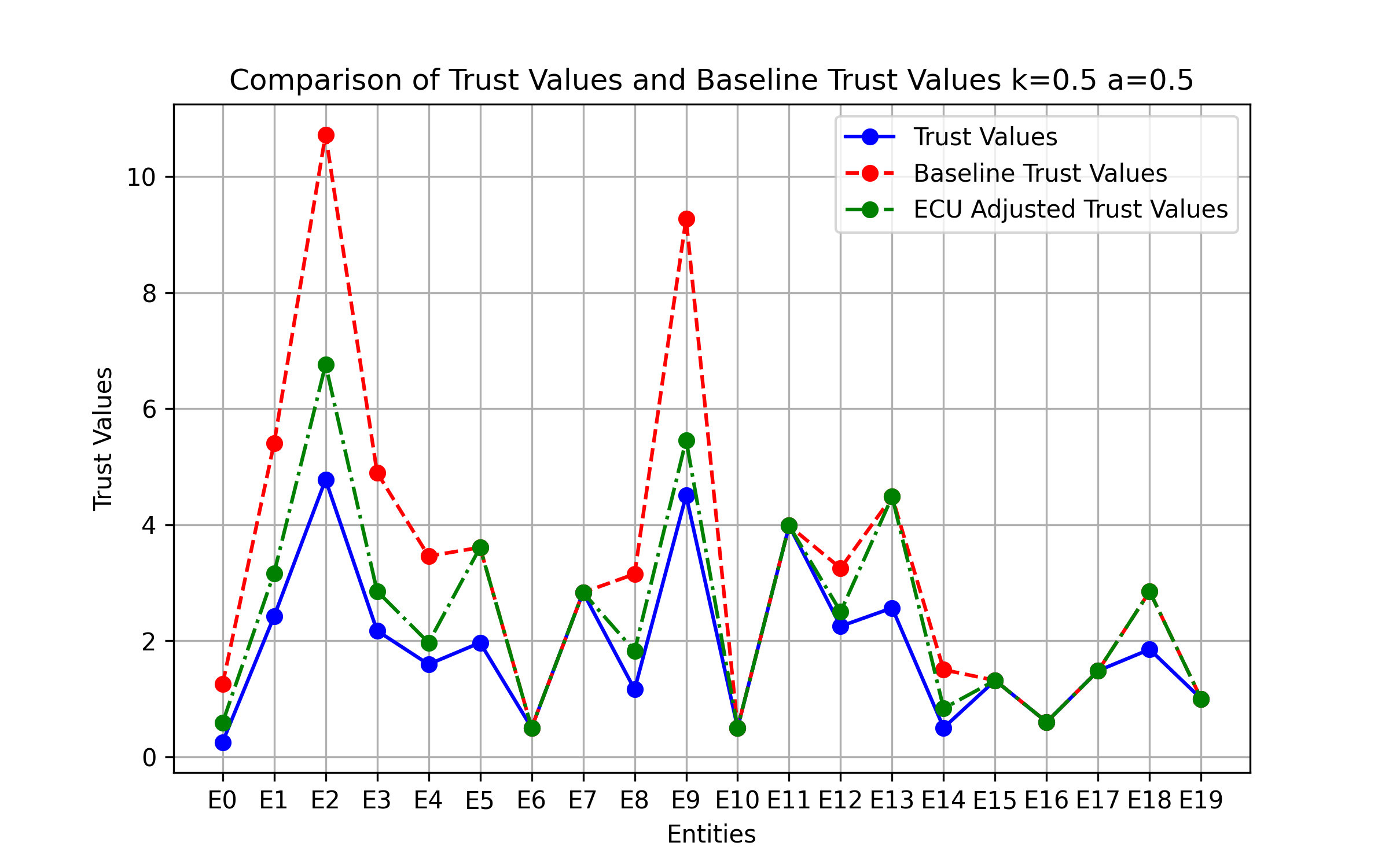}
  \caption{}
  \label{fig:5}
\end{subfigure} &
\begin{subfigure}{0.23\textwidth}
  \centering
  \includegraphics[width=\linewidth]{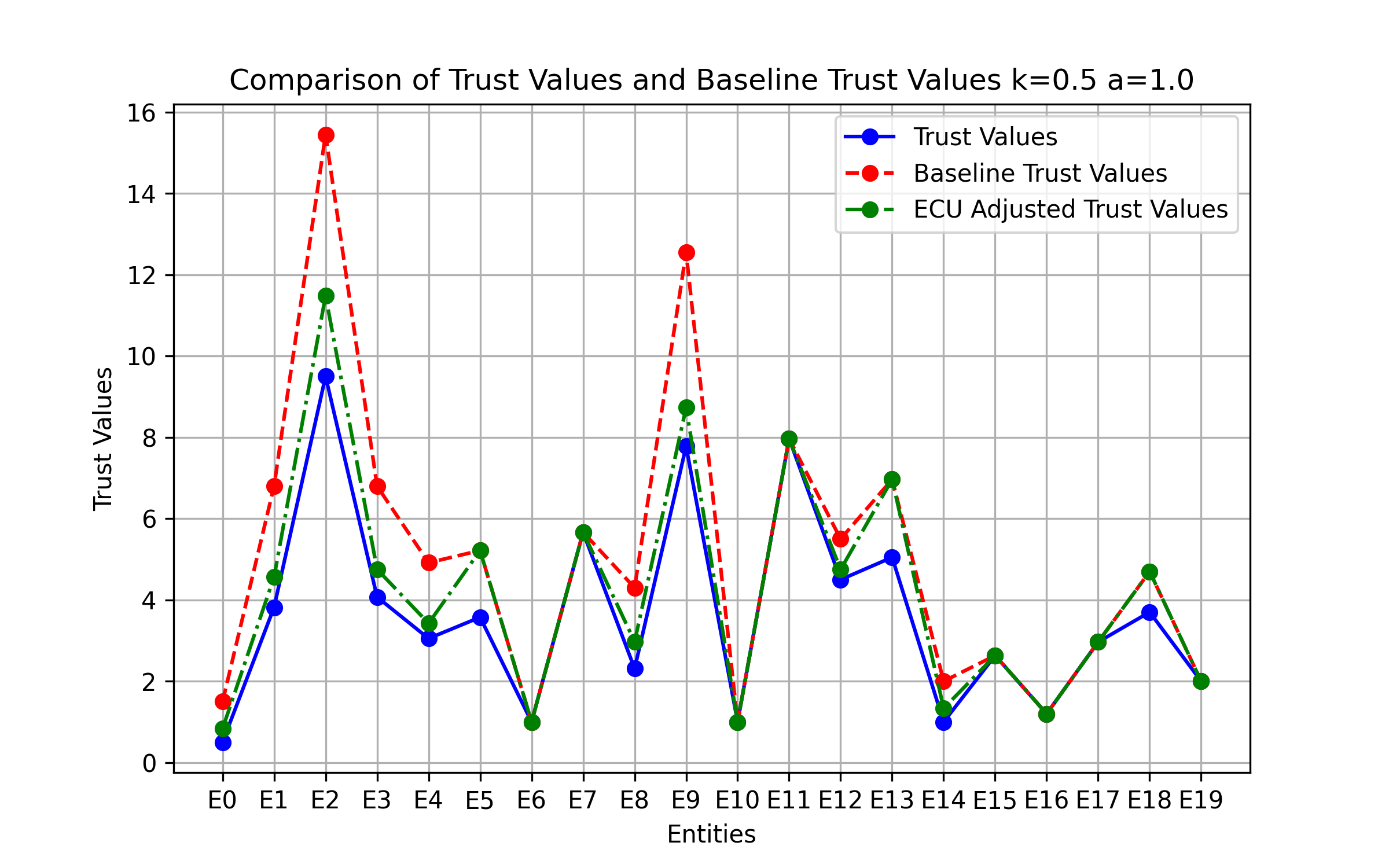}
  \caption{}
  \label{fig:6}
\end{subfigure} &
\begin{subfigure}{0.23\textwidth}
  \centering
  \includegraphics[width=\linewidth]{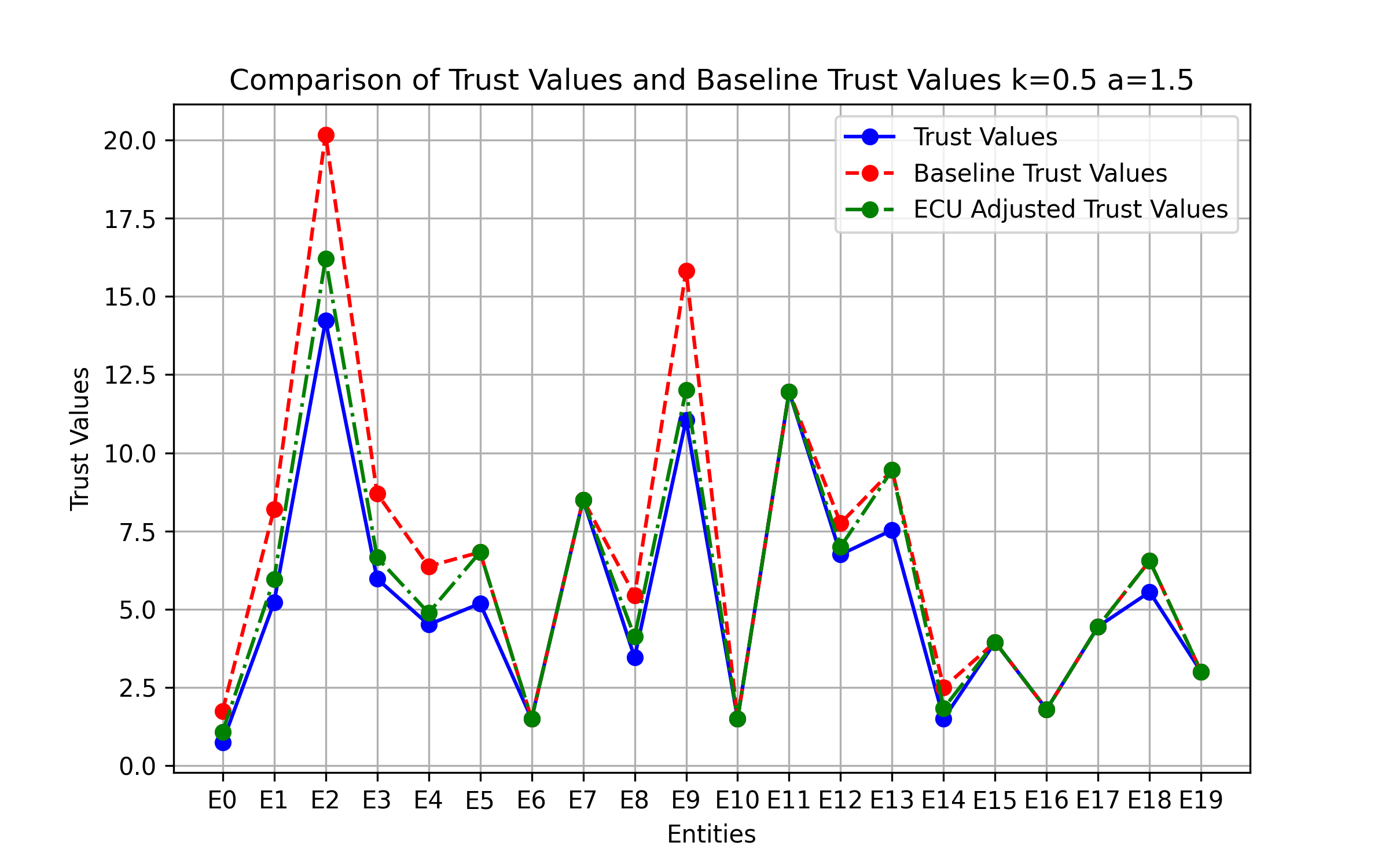}
  \caption{}
  \label{fig:7}
\end{subfigure} &
\begin{subfigure}{0.23\textwidth}
  \centering
  \includegraphics[width=\linewidth]{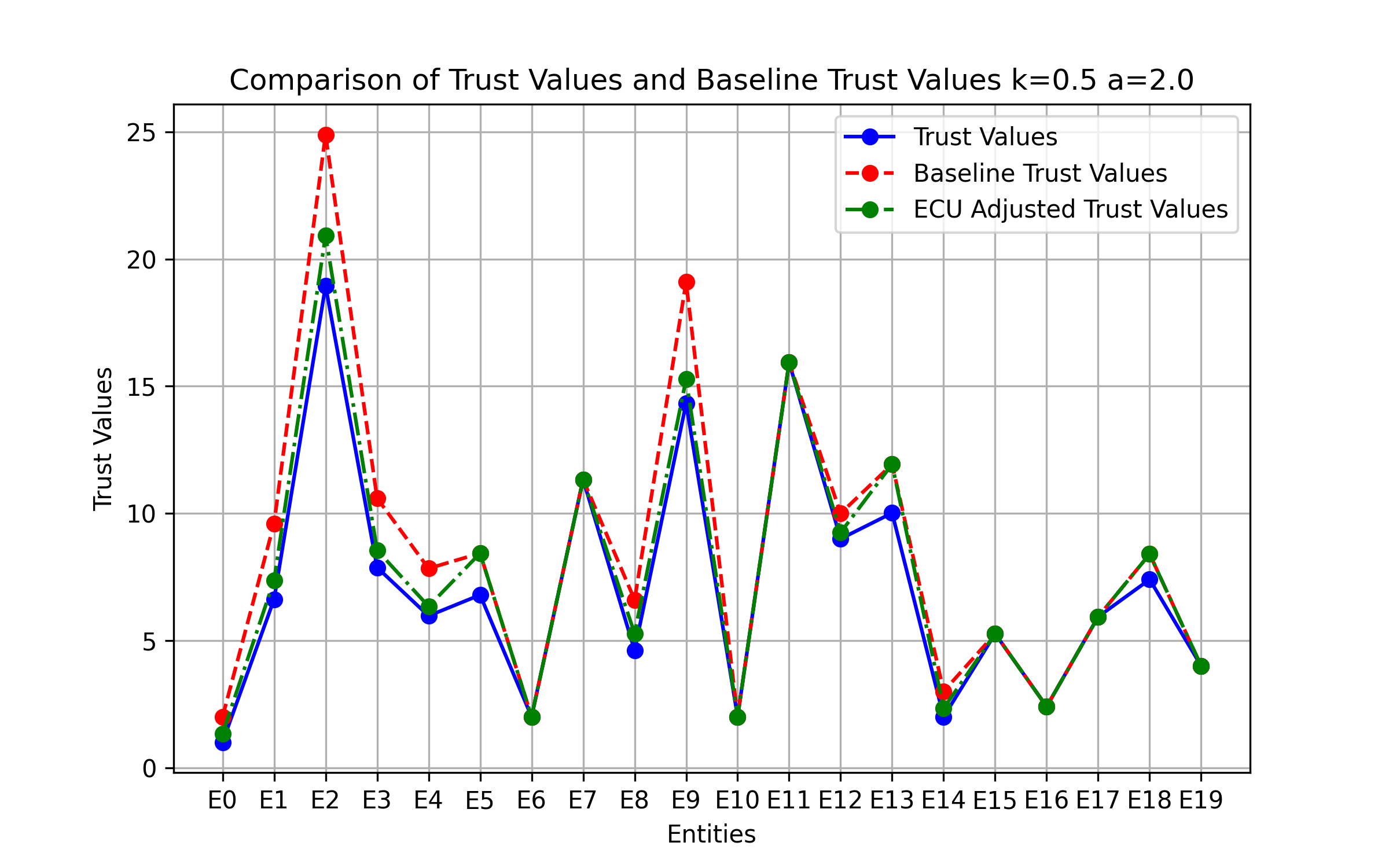}
  \caption{}
  \label{fig:8}
\end{subfigure} \\
\\
\begin{subfigure}{0.23\textwidth}
  \centering
  \includegraphics[width=\linewidth]{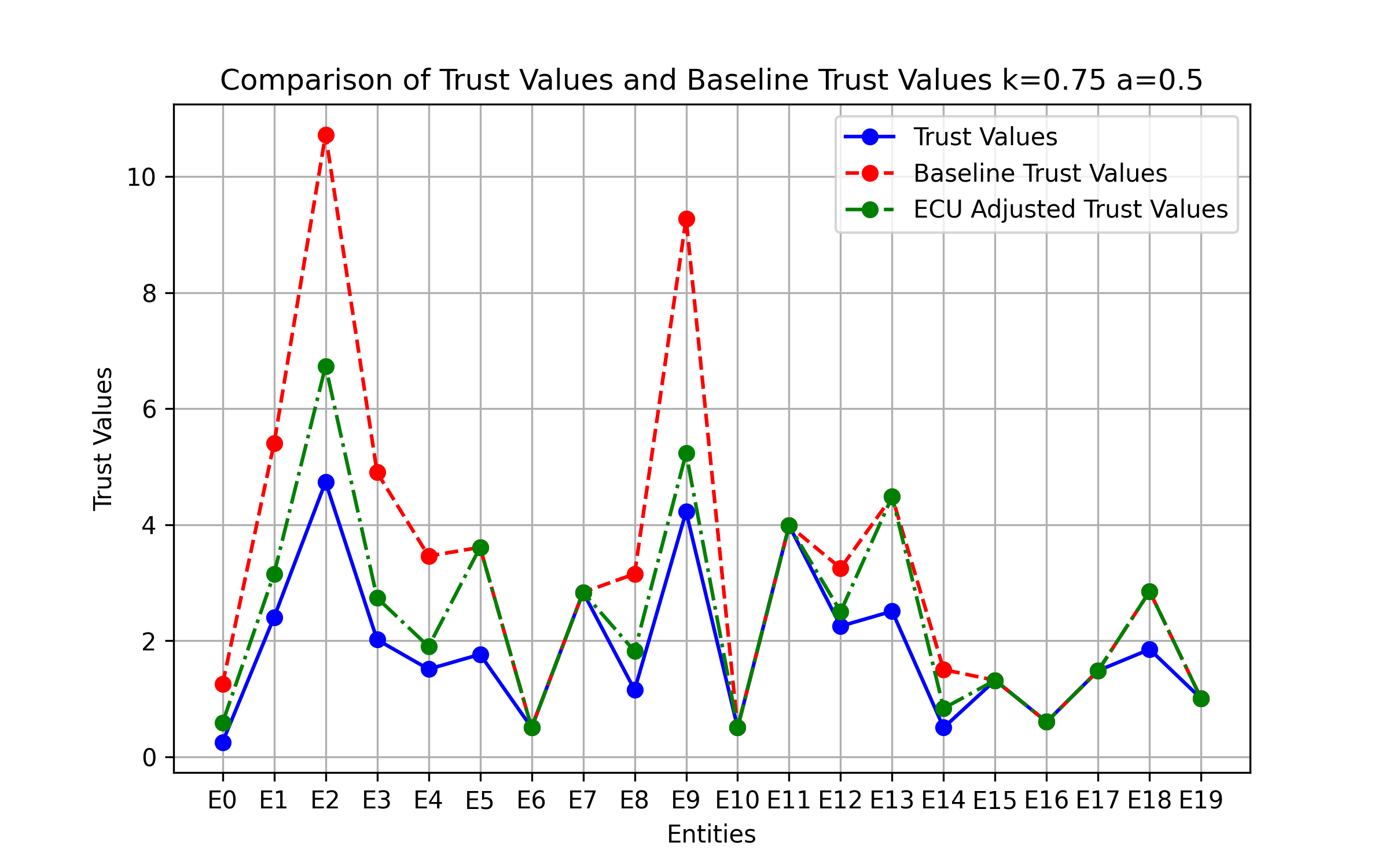}
  \caption{}
  \label{fig:9}
\end{subfigure} &
\begin{subfigure}{0.23\textwidth}
  \centering
  \includegraphics[width=\linewidth]{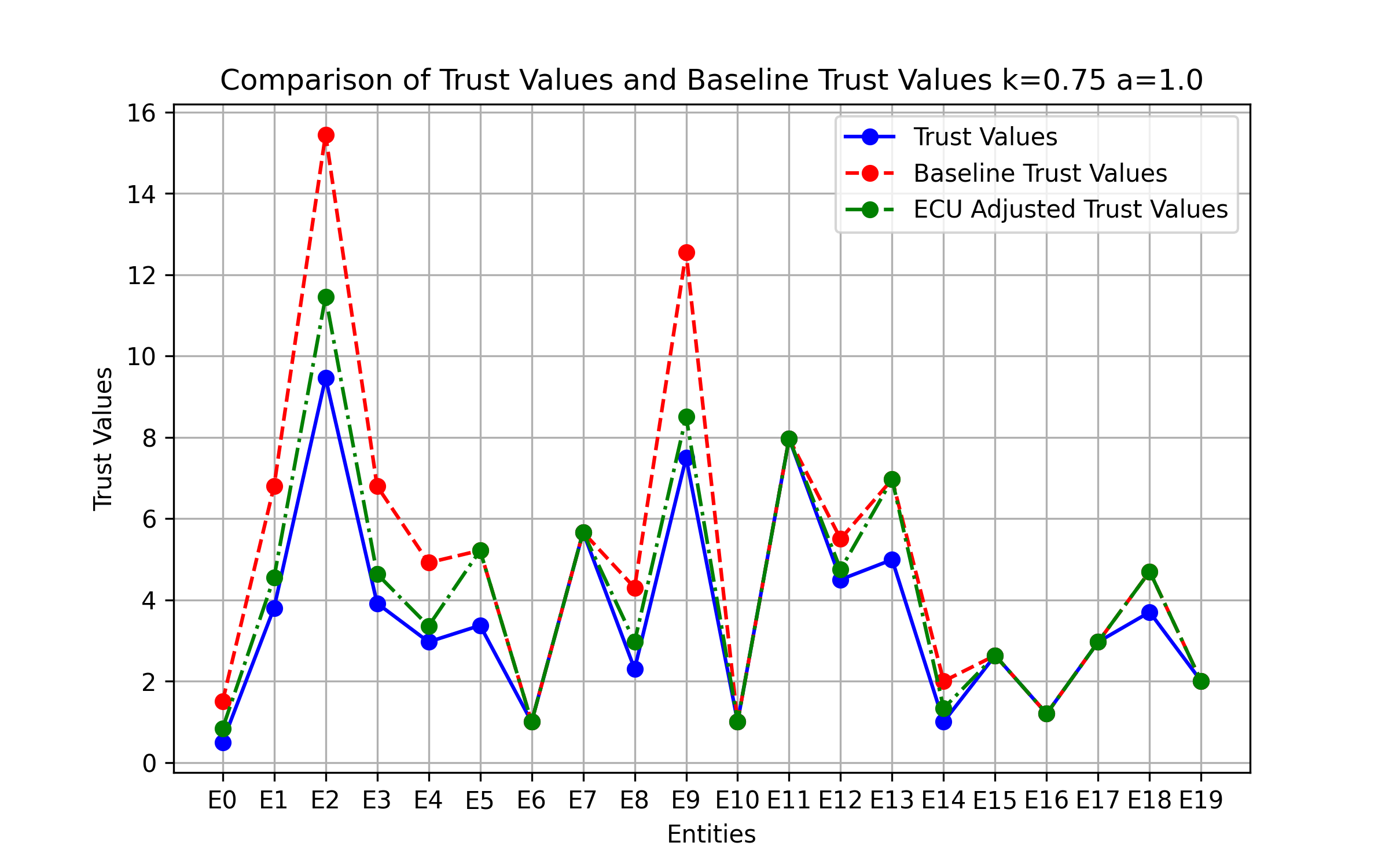}
  \caption{}
  \label{fig:10}
\end{subfigure} &
\begin{subfigure}{0.23\textwidth}
  \centering
  \includegraphics[width=\linewidth]{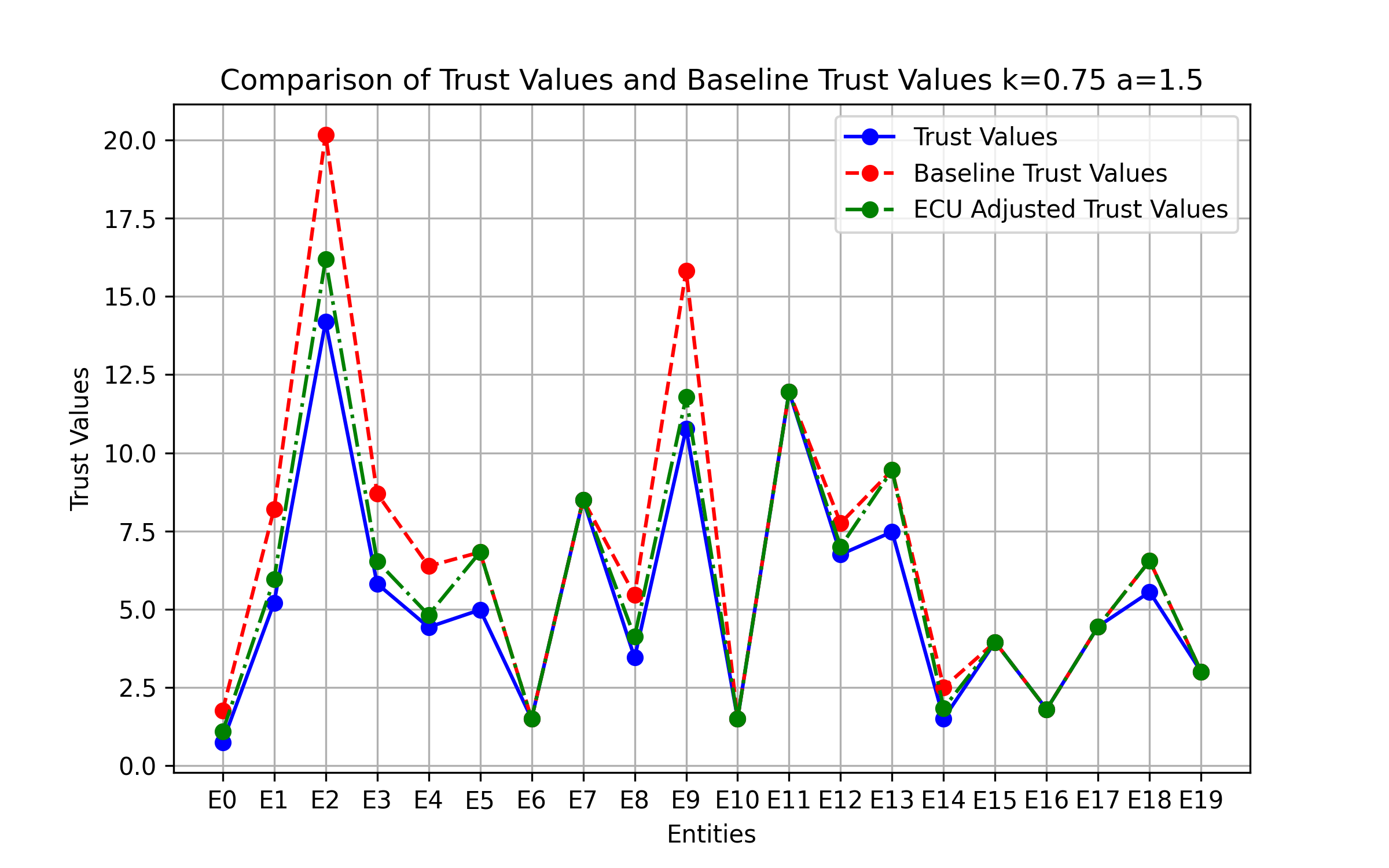}
  \caption{}
  \label{fig:11}
\end{subfigure} &
\begin{subfigure}{0.23\textwidth}
  \centering
  \includegraphics[width=\linewidth]{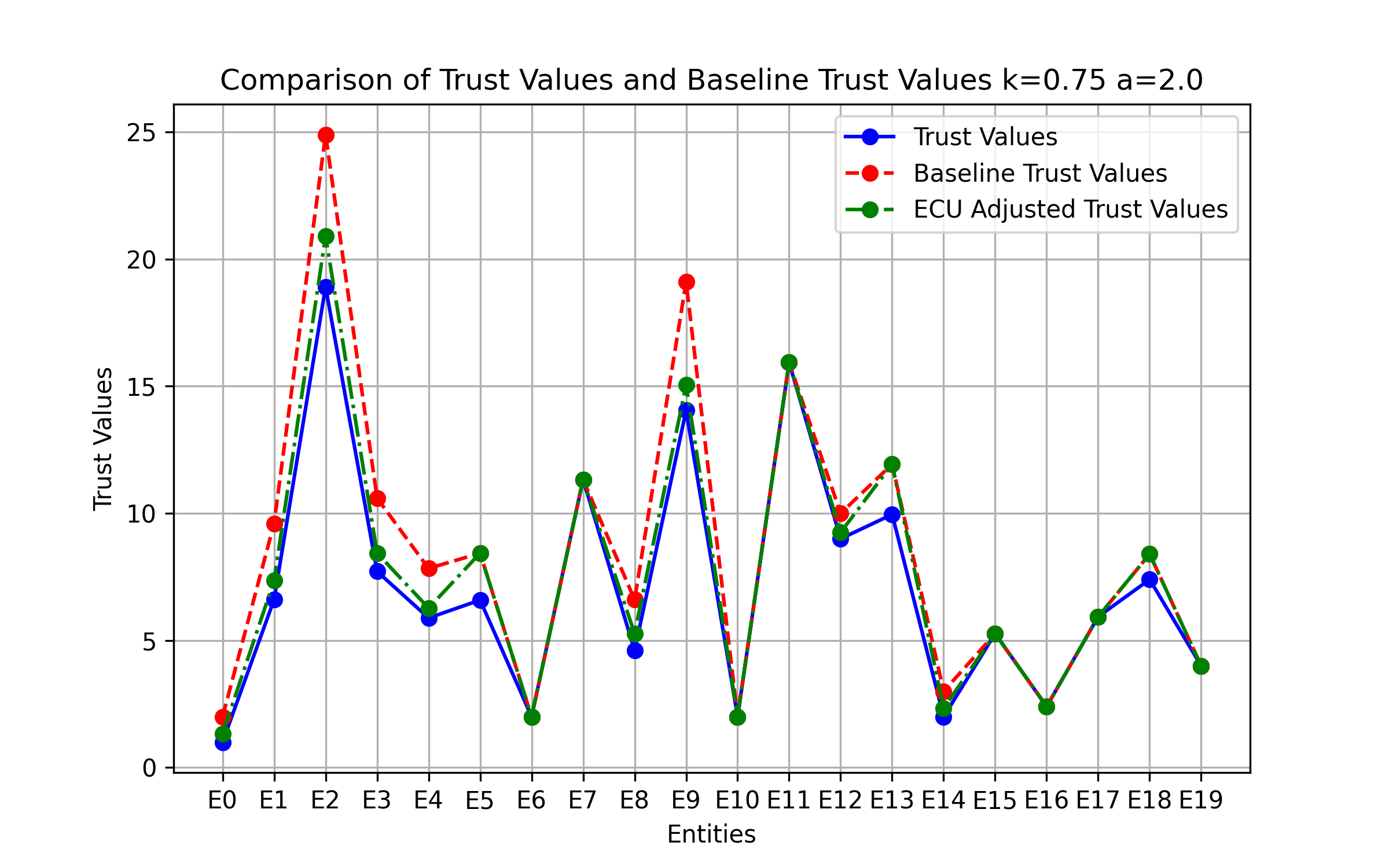}
  \caption{}
  \label{fig:12}
\end{subfigure} \\
\\
\begin{subfigure}{0.23\textwidth}
  \centering
  \includegraphics[width=\linewidth]{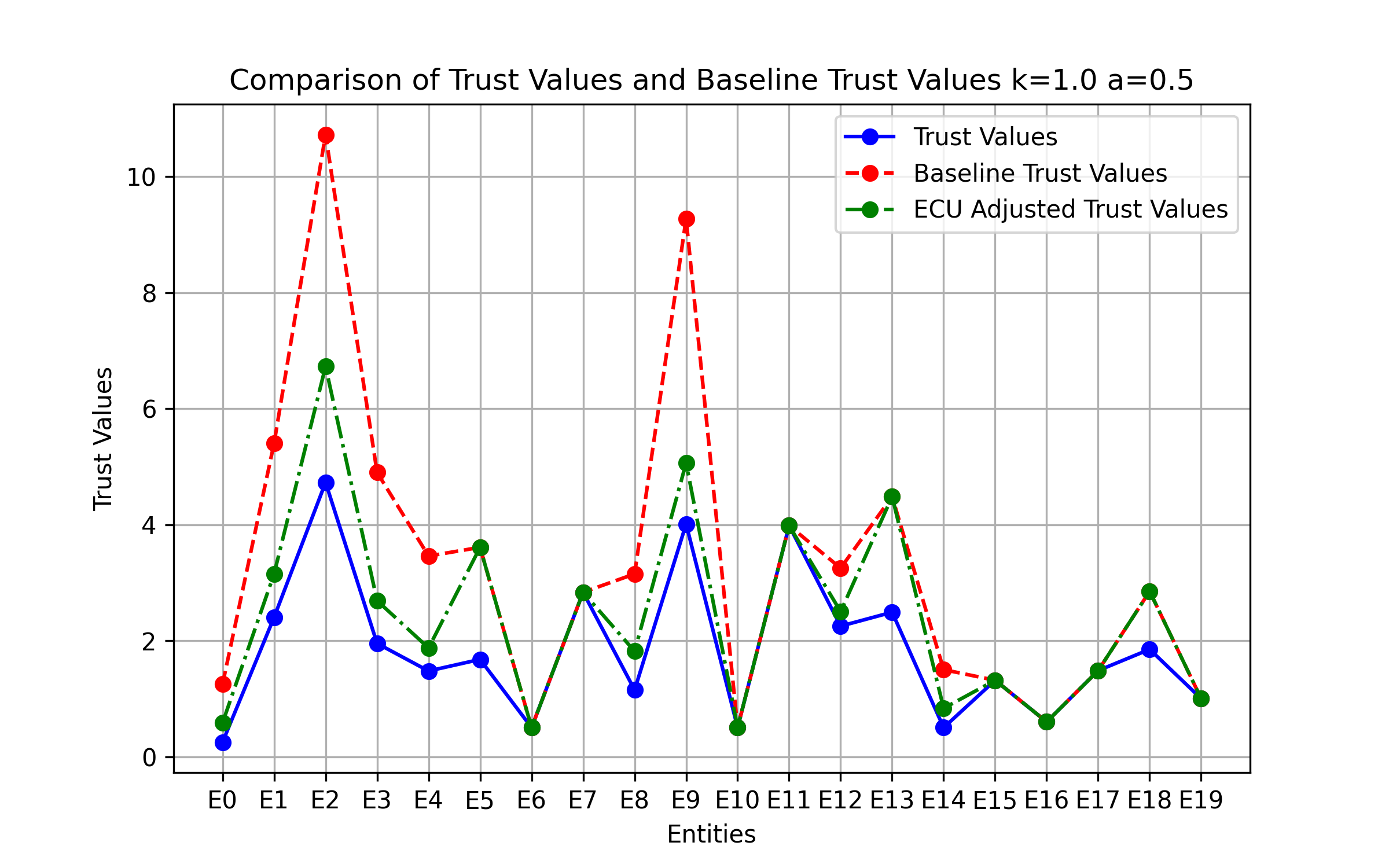}
  \caption{}
  \label{fig:13}
\end{subfigure} &
\begin{subfigure}{0.23\textwidth}
  \centering
  \includegraphics[width=\linewidth]{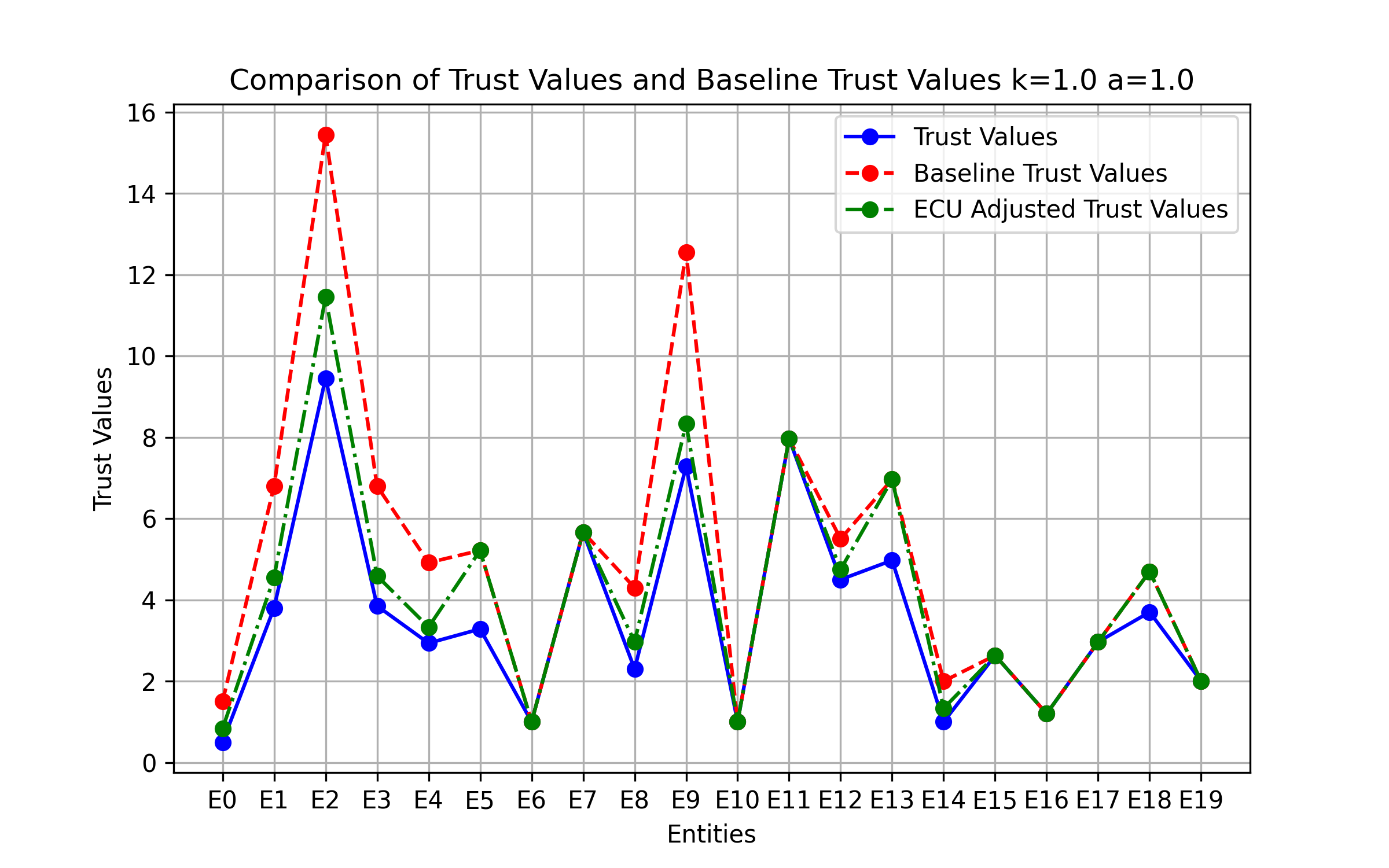}
  \caption{}
  \label{fig:14}
\end{subfigure} &
\begin{subfigure}{0.23\textwidth}
  \centering
  \includegraphics[width=\linewidth]{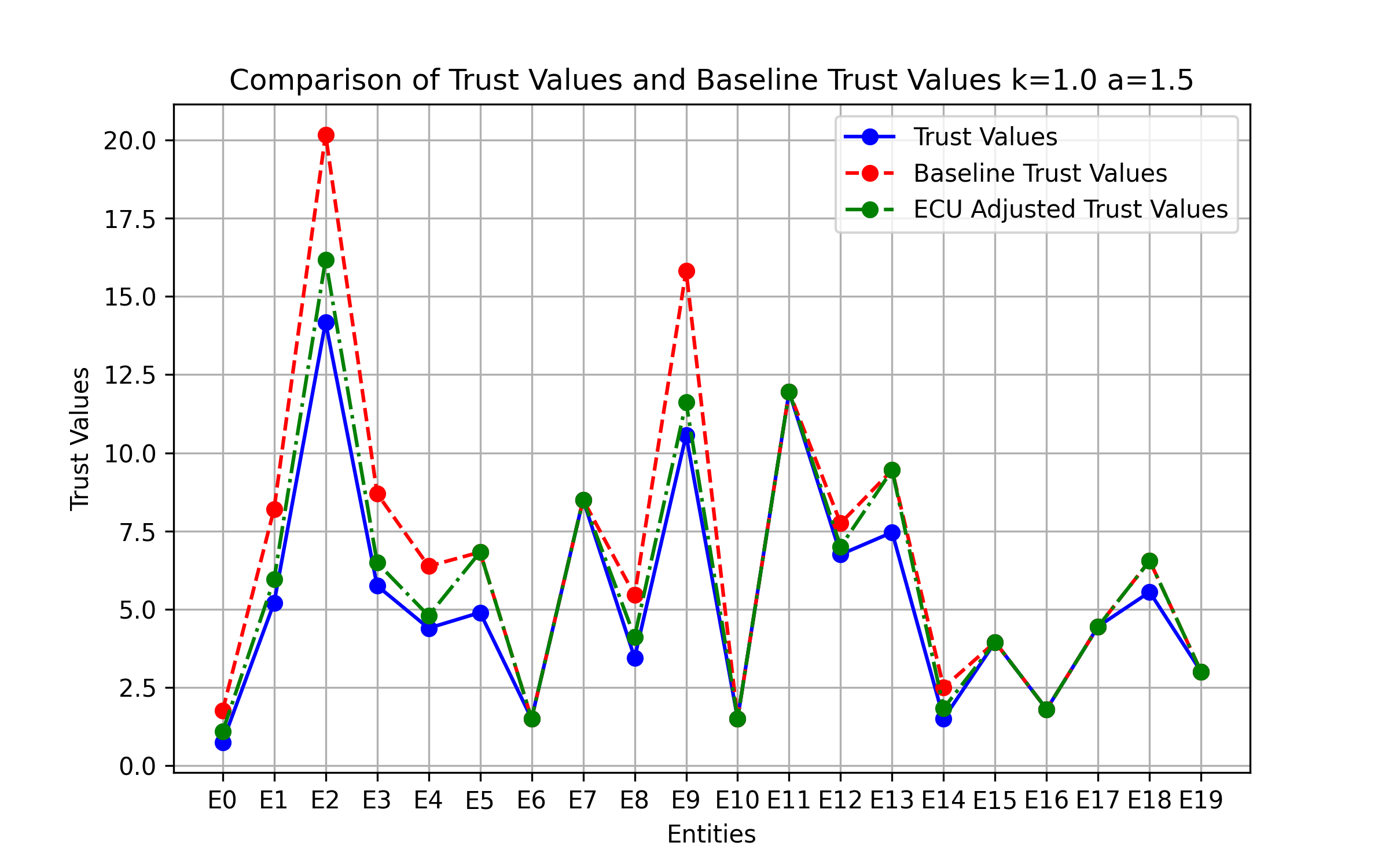}
  \caption{}
  \label{fig:15}
\end{subfigure} &
\begin{subfigure}{0.23\textwidth}
  \centering
  \includegraphics[width=\linewidth]{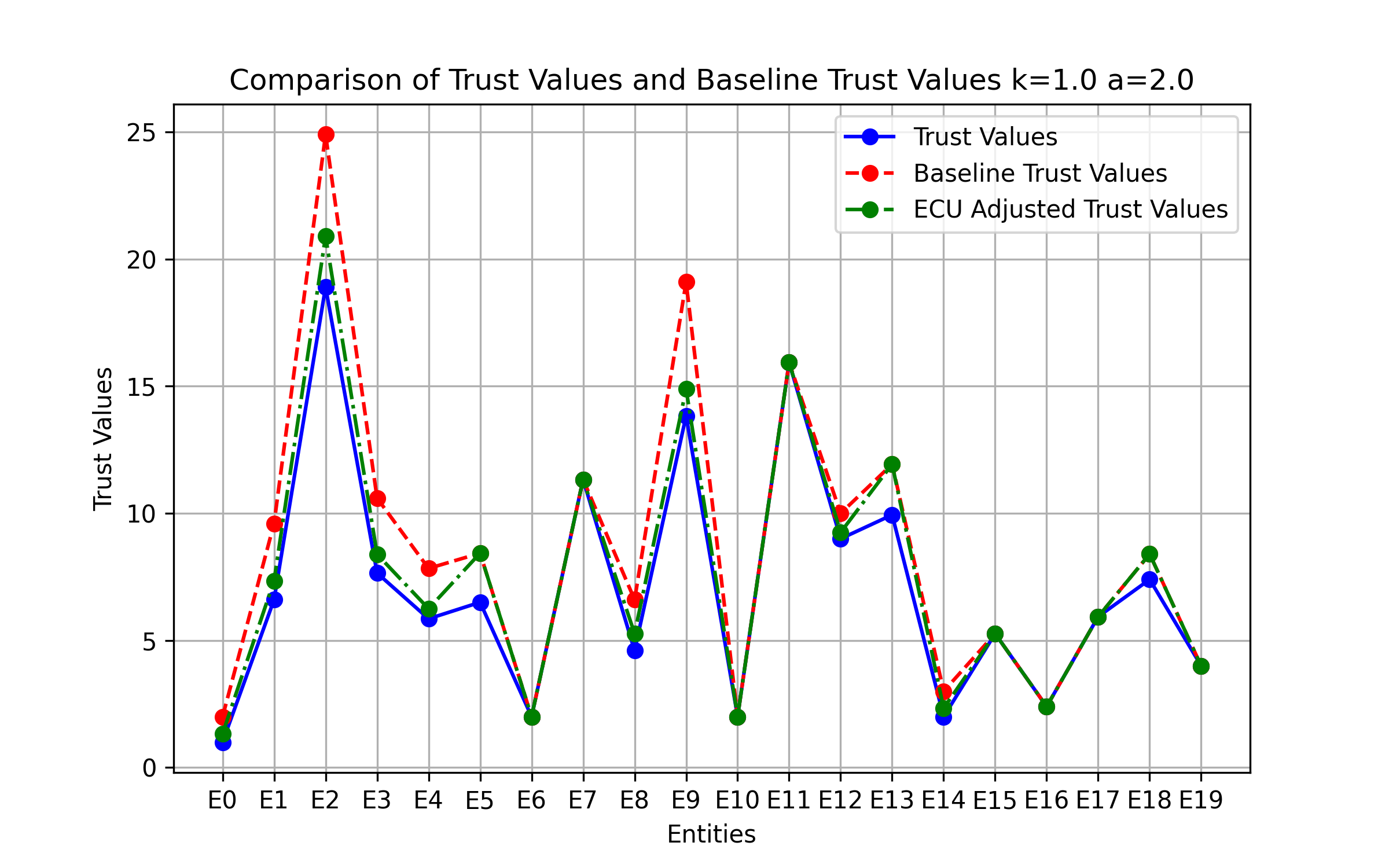}
  \caption{}
  \label{fig:16}
\end{subfigure} \\
\end{tabular}
\caption{Comparison of the  \emph{Baseline Trust Value}, \emph{Trust Value}, and \emph{ECU Adjusted Trust Value} under different combinations of the k and $alpha$ (denoted as a) values}
\label{Consolidated Result}
\end{figure*}
\section{Experimentation}
The performance of the proposed framework has been tested out under different scenarios using python. 
We utilized the NetworkX package \cite{hagberg2020networkx} to randomly generate the dependency graph of the in-vehicle network. The evaluation of the proposed framework was simulated using a graph with 20 nodes, where each node represented an ECU, and random connections were established as shown in Figure \ref{fig:ecu_invehicle}.

During our experiments with various value combinations, we established three key evaluation metrics: \emph{Baseline Trust Value}, \emph{Trust Value}, and \emph{ECU Adjusted Trust Value}. The \emph{Baseline Trust Value} for an $\mathcal{E\textsubscript{i}}$ indicates that there is no remote injection in the network and no discrepancy between the value observed from $\mathcal{E\textsubscript{i}}$ and the value calculated by $\mathcal{E\textsubscript{j}}$ for $\mathcal{E\textsubscript{i}}$ in any $\mathcal{E\textsubscript{i}} \to \mathcal{E\textsubscript{j}}$ connection. The \emph{Trust Value} for an $\mathcal{E\textsubscript{i}}$ is the sum of the values calculated by other $\mathcal{E\textsubscript{j}}$s, where an $\mathcal{E\textsubscript{i}} \to \mathcal{E\textsubscript{j}}$ connection exists in the in-vehicle network graph, as calculated in equation \ref{trust_ecu}. The \emph{ECU Adjusted Trust Value} for an $\mathcal{E\textsubscript{i}}$ represents the adjusted trust score, as shown in equation \ref{adjusted trust}, which is calculated after considering the feasibility of remote injection for an ECU, given by the parameter $\epsilon\textsubscript{i}$ (random distribution considered in our evaluation as shown in Fig. \ref{trust_ecu}), and the trust scores calculated in equation \ref{trust_ecu}.

\begin{equation}
EATV(\mathcal{E\textsubscript{i}}) = BTV(\mathcal{E\textsubscript{i}})-((BTV(\mathcal{E\textsubscript{i}})-\mathcal{T(E\textsubscript{i})})*(1-\epsilon\textsubscript{i}))
\label{adjusted trust}
\end{equation}

where $EATV,BTV$ is the \emph{Baseline Trust Value and ECU Adjusted Trust Value} of $\mathcal{E\textsubscript{i}}$ respectively.

The different combinations of the evaluation metrics talked about in the previous section along along with the impact of $k$ as mentioned in equation \ref{Eqn: Weight} and $\alpha$ (denoted as a) from equation \ref{trust_ecu} has been shown in Fig. \ref{Consolidated Result}. 

From Fig. \ref{attack difficulty}, it is evident that ECU, E5, E13, and E18 exhibit a low likelihood of being attacked, or in other words, they show greater resilience to remote injection. Additionally, E5, E13, and E18 possess relatively lower connectivity, with 3, 4, and 4 edges, respectively, as shown in the in-vehicle network in Figure \ref{fig:graph}. Due to this relatively low connectivity, the \emph{Baseline Trust Values} for these nodes, across all combinations of $k$ and $\alpha$ (represented as 'a' in Figure \ref{Consolidated Result}), remain relatively low. It is also observable across all the various combinations in Figure \ref{Consolidated Result} that the \emph{ECU Adjusted Trust Value} for E5, E13, and E18 is almost identical to their \emph{Baseline Trust Values}, despite the \emph{Trust Values} based on the other connected ECUs being lower. This outcome occurs because the ECUs connected to each of them exhibit a significantly higher likelihood of a remote injection attack, reducing the reliability of information from ECUs that are more vulnerable to injection attacks, compared to an ECU with a lower attack likelihood. As a result, by using eqn \ref{adjusted trust}, the \emph{ECU Adjusted Trust Value} remains relatively close to the \emph{Baseline Trust Values}. In contrast, the disparity between the \emph{Baseline Trust Values} and the \emph{ECU Adjusted Trust Value} is noticeably larger for E2 and E9 (as shown in Figure \ref{Consolidated Result}). This difference is attributed to the relatively higher chance of remote injection attacks on E2 and E9 (as indicated by Figure \ref{attack difficulty}). E2 is connected to E5, E17, E1, E11, E4, and E13, of which E5, E13, and E17 have a much lower chance of remote injection. Therefore, in the case of any discrepancies in the data calculations, the proposed framework can flag E2 as a potential source of attack, warranting further investigation to identify any additional vulnerabilities in the system. 
Thus, from the proposed framework it is to be noted that even the exposure of the different ECUs to the outside world play a significant role in determining whether there is a chance of a remote injection or not. If multiple vulnerable components are connected to a resilient component, as seen above in the example of E18, E13, and E5, it may not warrant an immediatiate investigation. This enhances the framework's resilience in scenarios where an attacker might intentionally try to deceive the system by fabricating a false scenario in which a resilient component appears to be under attack, while the actually compromised vulnerable components connected to the resilient one are overlooked. However, it is important to note that the system can lower the score of a resilient ECU if a sufficient number of ECUs contradict its readings, or if highly resilient ECUs contradict one another. In such cases, an Intrusion Detection System (IDS) can be designed to trigger an investigation. Such an IDS could be designed to trigger an investigation when the threshold or frequency of contradictory ECUs exceeds a certain level. The IDS would consider both the number of ECUs that contradict each other and the resilience of the contradicting ECUs to remote injection attacks, thereby enabling a more comprehensive evaluation of the data integrity from the entire in-vehicle network. 

It can also be observed from the results in Figure \ref{Consolidated Result} that the effect of the decay rate, $k$, in the trust calculation as defined in equation \ref{Eqn: Weight}, has a minimal impact on the trust values. This is because the value of $k$ is relatively small in the experiment, so any changes to it result in only a minor effect on the overall trust value. A potential solution to this limitation could be to consider a higher value for $k$ or to apply a logarithmic function to equation \ref{Eqn: Value inference} to amplify the distance, though such approaches fall outside the scope of this study. Additionally, the value of $\alpha$ (denoted as 'a' through subfigures a-p in Figure \ref{Consolidated Result}) significantly influences the \emph{Trust Value} for each ECU. This is due to the amplification of the remote injection probability, $\epsilon$, for each connected ECU in equation \ref{trust_ecu}, which in turn contributes to changes in the \ref{trust_ecu} values of all dependent ECUs.

\section{Conclusion}
In this paper, we have proposed a trust framework that assesses the trustworthiness of individual ECUs within an in-vehicle network, based on their connectivity to the outside world. By leveraging the interconnectivity of the ECUs within the in-vehicle network, the framework determines the vulnerability of an ECU to attack, considering the connectivity of the ECUs it is linked to. This framework is effective for analyzing and studying the scope and propagation of remote injection attacks, taking into account the ECUs' connectivity within the network. Additionally, the proposed framework can be utilized to design an efficient Intrusion Detection System (IDS) that does not rely on statistical data, the frequency of information received from the ECUs, or any data signatures that could be exploited by an attacker to replay or manipulate communication within the network. The framework can also help identify reliability issues within the in-vehicle system, even if an attacker exploits data patterns from specific ECUs or manipulates the frequency of injected data.





%
\bibliographystyle{unsrt}
\bibliography{references}
\thispagestyle{empty}

\end{document}